\documentclass [journal]{IEEEtran}
\usepackage{cite,graphicx,amsmath,amssymb,eufrak,mathrsfs}

\begin{document}
\title{Gaussian Relay Channel Capacity to \\Within a Fixed Number of Bits}
\author{Woohyuk~Chang,
        Sae-Young~Chung,
        and~Yong~H.~Lee
\thanks{
        Woohyuk Chang is with the Center for High-Performance Integrated Systems, KAIST,  Daejeon 305-701, Republic of Korea
        (e-mail: whchang@kaist.ac.kr).
        Sae-Young Chung and Yong H. Lee are with the Dept. of EE, KAIST, Daejeon 305-701, Republic of Korea
        (e-mail: \{sychung, yohlee\}@ee.kaist.ac.kr).}}
\maketitle


\newtheorem{theorem}{Theorem}
\newtheorem{lemma}{Lemma}
\newtheorem{corollary}{Corollary}

\begin{abstract}
In this paper, we show that the capacity of the three-node Gaussian relay channel can be
achieved to within 1 and 2 bit/sec/Hz using compress-and-forward and amplify-and-forward
relaying, respectively.

%

\end{abstract}

\section{Introduction}
Although both relay channels and interference channels are fundamental building blocks for constructing multiuser networks,
their single-letter capacity characterization has been open for decades.
In \cite{EtkinTseWang:06}, instead of struggling to find the exact capacity utilizing complicated achievable schemes,
the authors show that a simple Han--Kobayashi scheme can achieve the capacity of the two-user Gaussian interference channel
to within $1$ bit/sec/Hz for all values of the channel parameters.

In \cite{AvestimehrDiggaviTse:07}, this approach is further developed and generalized.
The authors first consider a deterministic channel model related to a given Gaussian channel and develop a scheme to achieve the capacity of such a deterministic
channel.
After getting an insight from the achievable scheme for the deterministic channel, they then develop a scheme to achieve the original Gaussian channel
capacity to within a constant number of bits for all values of the channel parameters.
This approach is called the {\it deterministic approach}.
As an example, they showed that the decode-and-forward (DF) relaying scheme originally proposed by Cover and El Gamal in \cite{CoverElGamal:79}
achieves the capacity of the three-node relay channel to within $1$ bit/sec/Hz
and a simple partial DF relaying scheme achieves the capacity of the diamond Gaussian relay channel to within $2$ bits/sec/Hz.

In this paper, we first show that
the compress-and-forward (CF) relaying scheme by Cover and El Gamal in \cite{CoverElGamal:79} can
achieve the three-node Gaussian relay channel capacity to within $1$ bit/sec/Hz for all values of the channel parameters.
We also show a simple amplify-and-forward (AF) relaying scheme where the relay amplifies and forwards the received signal
only when the channel from the source to the relay is stronger than the channel from the source to the destination
also achieves the Gaussian relay channel capacity to within $2$ bits/sec/Hz regardless of the channel
parameters.

The rest of the paper is organized as follows.
In Section \ref{Sec:ChannelModel}, we introduce the Gaussian relay channel.
Section \ref{Sec:OneBit} shows that the CF relaying scheme achieves the Gaussian relay channel capacity to within $1$ bit/sec/Hz
regardless of the channel parameters.
In Section \ref{Sec:TwoBits}, the AF relaying scheme is proposed and shown to achieve the capacity to within $2$ bits/sec/Hz.


\section{Gaussian Relay Channel} \label{Sec:ChannelModel}
\begin{figure}[h]
\centering
\includegraphics[width=2.6in]{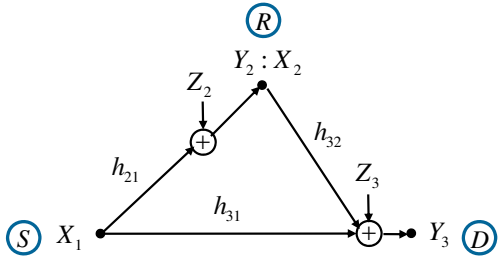}
\caption{Gaussian relay channel.}
\label{Fig:RelayChannel}
\end{figure}

We consider a Gaussian relay channel in Fig. \ref{Fig:RelayChannel}.
For simplicity, we assume a full-duplex relay as in \cite{AvestimehrDiggaviTse:07,CoverElGamal:79,KramerGastparGupta:05}.
The received signals at the relay and the destination are given by
\setlength{\arraycolsep}{0.0em}
\begin{eqnarray}
&&Y_2=h_{21}X_1+Z_2,  \\
&&Y_3=h_{31}X_1+h_{32}X_2+Z_3,
\end{eqnarray}
respectively, where $h_{21}$, $h_{31}$, and $h_{32}$ are complex constants,
$Z_2\sim\mathcal{CN}(0,1)$ and $Z_3\sim\mathcal{CN}(0,1)$ are noises at the relay and at the destination, respectively,
that are independent of each other,
$\mathbb E[|X_1|^2]\leq P_1$, and $\mathbb E[|X_2|^2]\leq P_2$.

\begin{figure}[t]
\centering
\includegraphics[width=2.9in]{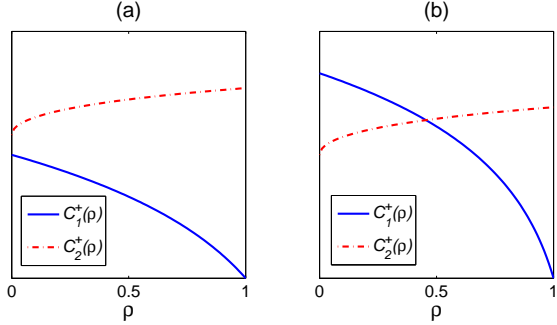}
\caption{$C_1^+(\rho)$ and $C_2^+(\rho)$: (a) $|h_{21}|^2P_1 \leq |h_{32}|^2P_2$, (b) $|h_{21}^2|P_1 > |h_{32}^2|P_2$.}
\label{Fig:UB}
\end{figure}

From \cite{CoverElGamal:79}, the upper bound on its channel capacity can be found as
\setlength{\arraycolsep}{0.0em}
\begin{eqnarray}
&&C^+ = \max_{0\leq \rho\leq 1}\min\left\{ C_1^+(\rho), C_2^+(\rho) \right\}
\label{Eqn:UB}
\end{eqnarray}
where
\setlength{\arraycolsep}{0.0em}
\begin{eqnarray}
C_1^+(\rho) &{}={}& \log_2\left( 1+(1-\rho^2)\left( |h_{21}|^2+|h_{31}|^2 \right)P_1\right), \label{Eqn:C_1^+}\\
C_2^+(\rho) &{}={}& \log_2\Big(1+|h_{31}|^2P_1+|h_{32}|^2P_2    \nonumber \\
&&        \qquad\quad+2\rho\sqrt{|h_{31}|^2|h_{32}|^2P_1P_2} \Big), \label{Eqn:C_2^+} \\
\rho &{}={}&\frac{\mathbb{E}\left[X_1 X_2^\dag\right]}{\sqrt{\mathbb{E}\left[|X_1|^2\big]\mathbb{E}[|X_2|^2\right]}},
\end{eqnarray}
where $X_2^\dag$ is the complex conjugate of $X_2$.
Since $C_1^+(\rho)$ decreases while $C_2^+(\rho)$ increases as $\rho$ increases, and $C_1^+(1)\leq C_2^+(1)$,
there exist two possible cases for the optimal $\rho^*$ values depending on whether $|h_{21}|^2P_1 \leq |h_{32}|^2P_2$ or not as shown in Fig \ref{Fig:UB}.
\begin{itemize}
\item If $|h_{21}|^2P_1 \leq |h_{32}|^2P_2$, $C_1^+(\rho)\leq C_2^+(\rho)$ for all $0\leq \rho\leq 1$, and hence $\rho^*=0$ and $C^+=C_1^+(0)$.
\item If $|h_{21}|^2P_1 > |h_{32}|^2P_2$, $\rho^*$ is determined such that $C^+=C_1^+(\rho^*)=C_2^+(\rho^*)$.
\end{itemize}

\section{Gaussian Relay Channel Capacity to Within One Bit: CF Relaying Scheme} \label{Sec:OneBit}

\begin{figure}
\centering
\includegraphics[width=2.9in]{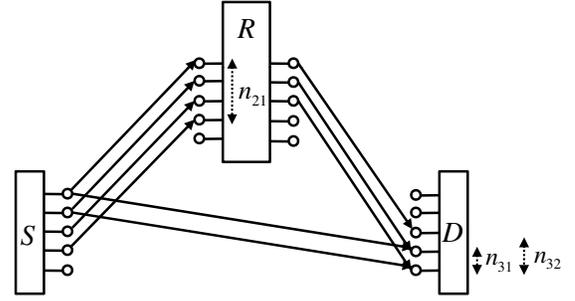}
\caption{Deterministic relay channel.}
\label{Fig:DeterministicRelayChannel}
\end{figure}

In \cite{AvestimehrDiggaviTse:07}, the authors introduce a deterministic relay channel model corresponding to the Gaussian relay channel
in Section \ref{Sec:ChannelModel} as shown in Fig. \ref{Fig:DeterministicRelayChannel}.
Each circle at the transmitter of each node represents a signal level and a binary digit can be put on each circle for transmission.
$n_i$ represents the received signal-to-noise ratio (SNR) for path $i$ in dB scale.
More specifically, $n_{31}=\lceil \log_2 (h_{31}P_1)\rceil$, $n_{32}=\lceil \log_2 (h_{32}P_2)\rceil$ and $n_{21}=\lceil \log_2 (h_{21}P_1)\rceil$.
Then, the transmitted bits at the first $n_i$ signal levels are received clearly through the path $i$ at the corresponding receiver
while the remaining bits at the other signal levels are not delivered to the receiver through the path $i$.
This is motivated to mimic the AWGN channel since the effect of background Gaussian noise can be simplified
such that the first $n_i$ bits including the most significant bit (MSB) are above noise level at the receiver while
the remaining bits including the least significant bit (LSB) are below noise level.

The capacity of this deterministic relay channel is then found in \cite{AvestimehrDiggaviTse:07} as
\setlength{\arraycolsep}{0.0em}
\begin{eqnarray}
C_d &{}={}& \min \left\{ \max\{n_{21},n_{31}\},\max\{n_{32},n_{31}\} \right\} \nonumber \\
&{}={}& n_{31} + \left[\min\{n_{21},n_{32}\}-n_{31} \right]^+
\label{Eqn:DRelayChCapacity}
\end{eqnarray}
where $[x]^+=\max\{x,0\}$.
(\ref{Eqn:DRelayChCapacity}) implies a capacity-achieving scheme such that
first $n_{31}$ bits are directly delivered to the destination from the source while the remaining $\left[\min\{n_{21},n_{31}\}-n_{31} \right]^+$ bits
are routed from the source to the destination through the relay.
This motivates the authors in \cite{AvestimehrDiggaviTse:07} to propose a DF-based relaying scheme such that 
it chooses one of two schemes depending on whether $|h_{31}|^2 > |h_{21}|^2$ or not as follows.
\begin{itemize}
\item If $|h_{31}|^2 > |h_{21}|^2$, the relay is ignored and the achievable rate is equal to $\log_2(1+|h_{31}|^2P_1)$.
\item If $|h_{31}|^2 \leq |h_{21}|^2$, the block-Markov encoded DF scheme in \cite{CoverElGamal:79} is used and hence its achievable rate is
equal to $\min\{\log_2(1+|h_{21}|^2P_1), \log_2(1+|h_{31}|^2P_1+|h_{32}|^2P_2) \}$.
\end{itemize}
Hence, the overall achievable rate is given by
\begin{eqnarray}
&&R_{DF}= \max\big\{ \log_2(1+|h_{31}|^2P_1),
\min\{\log_2(1+|h_{21}|^2P_1), \nonumber \\
&&\qquad\qquad \log_2(1+|h_{31}|^2P_1+|h_{32}|^2P_2) \} \big\},
\label{Eqn:R_DF}
\end{eqnarray}
and $C^+-R_{DF}\leq 1$ is shown to be satisfied for all values of the channel parameters.

In \cite{Host-MadsenZhang:05,KramerGastparGupta:05,KramerMaricYates:07},
the achievable rate of the CF relaying scheme for the Gaussian relay channel in Section \ref{Sec:ChannelModel} is explicitly given by
\setlength{\arraycolsep}{0.0em}
\begin{eqnarray}
&&R_{CF}= \nonumber \\
&&\;\;\log_2\left(1+|h_{31}|^2P_1+\frac{|h_{21}|^2|h_{32}|^2P_1P_2}{1+\left(|h_{31}|^2+|h_{21}|^2\right)P_1+|h_{32}|^2P_2} \right). \nonumber \\
\label{Eqn:OriginalR_CF}
\end{eqnarray}
Then, $R_{CF}$ in (\ref{Eqn:OriginalR_CF}) can be rewritten as
\setlength{\arraycolsep}{0.0em}
\begin{eqnarray}
R_{CF}&{}={}& \log_2\left(1+|h_{31}|^2P_1\right) \nonumber \\
&&+ \log_2\left(1+\delta\cdot\frac{\min\left\{ |h_{21}|^2P_1, |h_{32}|^2P_2 \right\}}{1+|h_{31}|^2P_1} \right),
\label{Eqn:ModifiedR_CF}
\end{eqnarray}
where
\setlength{\arraycolsep}{0.0em}
\begin{eqnarray}
\delta&{}={}& \frac{\max\left\{ |h_{21}|^2P_1, |h_{32}|^2P_2 \right\}}{1+\left(|h_{31}|^2+|h_{21}|^2\right)P_1+|h_{32}|^2P_2}<1,
\end{eqnarray}
and $\delta\rightarrow 1$ as $\max\left\{ |h_{21}|^2P_1, |h_{32}|^2P_2 \right\}\rightarrow\infty$.
Interestingly, we can see that (\ref{Eqn:ModifiedR_CF}) is very similar to (\ref{Eqn:DRelayChCapacity}) such that
$\log_2\left(1+|h_{31}|^2P_1\right)$ bits are achieved from the direct path between the source and the destination
while $\log_2\left(1+\delta\cdot\min\left\{ |h_{21}|^2P_1, |h_{32}|^2P_2 \right\}/\left(1+|h_{31}|^2P_1\right) \right)$ bits are additionally achieved
through the relaying path.
This makes us conjecture that $C^+-R_{CF}\leq 1$ is also satisfied for all values of the channel parameters.

In this paper, we show that $C^+-R_{CF}\leq 1$ is indeed satisfied for all values of the channel parameters.
For simplicity, we define $a\triangleq |h_{31}|^2P_1$, $b\triangleq |h_{32}|^2P_2$, and $c\triangleq |h_{21}|^2P_1$.
We first consider the case of $c\leq b$ and then the case of $b<c$.

\subsection{The case of $c\leq b$ $\left(|h_{21}|^2P_1\leq |h_{32}|^2P_2\right)$}
Since $\frac{bc}{1+a+b+c}$ increases as $b$ increases,
\setlength{\arraycolsep}{0.0em}
\begin{eqnarray}
&&R_{CF}=\log_2\left(1+a+\frac{bc}{1+a+b+c} \right)   \nonumber \\
&&\quad\geq \log_2\left(1+a+\frac{c^2}{1+a+2c}\right)=\log_2\frac{(1+a+c)^2}{1+a+2c}.\;\; \label{Eqn:LBofRcf}
\end{eqnarray}
Defining $\Delta_1\triangleq C^+ -R_{CF}$, we get
\setlength{\arraycolsep}{0.0em}
\begin{eqnarray}
\Delta_1 &{}\leq{}& C^+ -\log_2\frac{(1+a+c)^2}{1+a+2c}     \nonumber \\
&{}={}& \log_2\left(1+a+c\right)-\log_2\frac{(1+a+c)^2}{1+a+2c}     \nonumber \\
&{}={}& \log_2\frac{1+a+2c}{1+a+c} \leq \log_2 2=1.
\end{eqnarray}
Hence, the CF relaying scheme achieves the Gaussian relay channel capacity to within one bit when $c\leq b$.
Moreover, when $c=b\rightarrow\infty$, we get $\Delta_1\rightarrow 1$.
On the other hand, when $b\rightarrow\infty$ only, $\Delta_1\rightarrow 0$, i.e., the capacity is asymptotically achieved.

\subsection{The case of $b<c$ $\left(|h_{32}|^2P_2 < |h_{21}|^2P_1\right)$}
Since $\rho^*$ in $C^+$ is determined to satisfy $C^+=C_1^+(\rho^*)=C_2^+(\rho^*)$,
\setlength{\arraycolsep}{0.0em}
\begin{eqnarray}
&&\quad\log_2\left( 1+(1-\rho^{*2})(a+c) \right)= \log_2\left(1+a+b+2\rho^*\sqrt{ab} \right)  \nonumber \\
&&\Longleftrightarrow (a+c)\rho^{*2} + 2\sqrt{ab}\rho^*+b-c = 0     \nonumber \\
&&\Longleftrightarrow \rho^* = \frac{\sqrt{(a-b+c)c}-\sqrt{ab}}{a+c},
\end{eqnarray}
where $ab < (a-b+c)c$ for $b<c$.
Hence,
\setlength{\arraycolsep}{0.0em}
\begin{eqnarray}
C^+ = \log_2 \left(1+a+b+\frac{2\sqrt{abc(a-b+c)}-2ab}{a+c} \right). \quad
\label{Eqn:UB-Case2}
\end{eqnarray}
Then,
\setlength{\arraycolsep}{0.0em}
\begin{eqnarray}
\Delta_1
&{}={}& \log_2\frac{(1+a+b+c)\left(1+a+b+\frac{2\sqrt{abc(a-b+c)}-2ab}{a+c} \right)}{(1+a)(1+a+b+c)+bc}     \nonumber \\
&{}={}& \log_2\left(1 + \frac{B+\sqrt{C}}{A}  \right),
\end{eqnarray}
where
\setlength{\arraycolsep}{0.0em}
\begin{eqnarray}
A &{}={}& ((1+a)(1+a+b+c)+bc)(a+c),    \\
B &{}={}& bc-ab-abc-a^2b+b^2c-ab^2,   \\
C &{}={}& 4(1+a+b+c)^2abc(a-b+c).
\end{eqnarray}
Note that if $\frac{B+\sqrt{C}}{A} \leq 1$, $\Delta_1 \leq 1$.
Since $A\geq B$ and $C\geq 0$, showing $(A-B)^2-C \geq 0$ is equivalent to showing $\frac{B+\sqrt{C}}{A} \leq 1$.
After some manipulation, we obtain
\begin{eqnarray}
(A-B)^2-C = (a+c)^2 \left( \alpha_0 + 2\alpha_1 c + \alpha_2 c^2 + \frac{\alpha_3}{a+c} \right), \nonumber \\
\end{eqnarray}
where
\begin{eqnarray}
\alpha_0 &{}={}& a^4+4a^3(b+1)+6a^2(b+1)^2+(b^2-1)^2 \nonumber \\
&&\quad +4a(b^3+b^2+b+1),     \label{Eqn:alpha_0}  \\
\alpha_1 &{}={}& (a+1)^3+(a^2+1)b-(a+1)b^2-b^3,       \label{alpha_1} \\
\alpha_2 &{}={}& (a-b)^2+2(a+b)+1,    \label{Eqn:alpha_2}    \\
\alpha_3 &{}={}& 4(ab+2a^2b+2ab^2+2a^2b^2+ab^3).
\end{eqnarray}
Since $a+c>0$ and $\alpha_3\geq 0$, it is sufficient to show $f(c)\triangleq\alpha_0 + 2\alpha_1 c + \alpha_2 c^2 \geq 0$ for all $c \geq 0$.
After some manipulation, we obtain the discriminant $D$ of $f(c)$ as
\begin{eqnarray}
D &{}={}& \alpha_1^2-\alpha_0\alpha_2  \nonumber \\
  &{}={}& -4ab\big\{ 2a^3+(3b+5)a^2+(8b^2+11b+4)a   \nonumber \\
  &&+(b+1)^2(3b+1) \big\} \leq 0,
\end{eqnarray}
which implies that $f(c)\geq 0$ for all $c$ since $\alpha_2 >0$.
Hence, $\Delta_1 \leq 1$ when $c>b$.
Finally, we conclude the CF relaying scheme achieves the Gaussian relay channel capacity to within $1$ bit/sec/Hz for all values of the channel parameters.

\begin{figure}[h]
\centering
\includegraphics[width=3.5in]{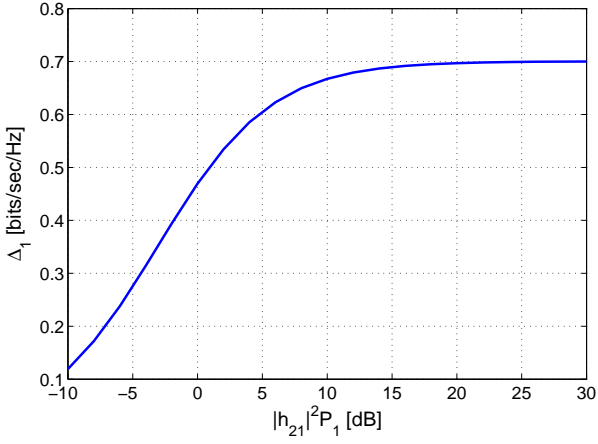}
\caption{$\Delta_1$ for various $|h_{21}|^2P_1$ values with $P_1=P_2$, $|h_{31}|^2=0.1|h_{21}|^2$ and $|h_{32}|^2=1.5|h_{21}|^2$.}
\label{Fig:Case1Results_CF}
\end{figure}

\begin{figure}[h]
\centering
\includegraphics[width=3.5in]{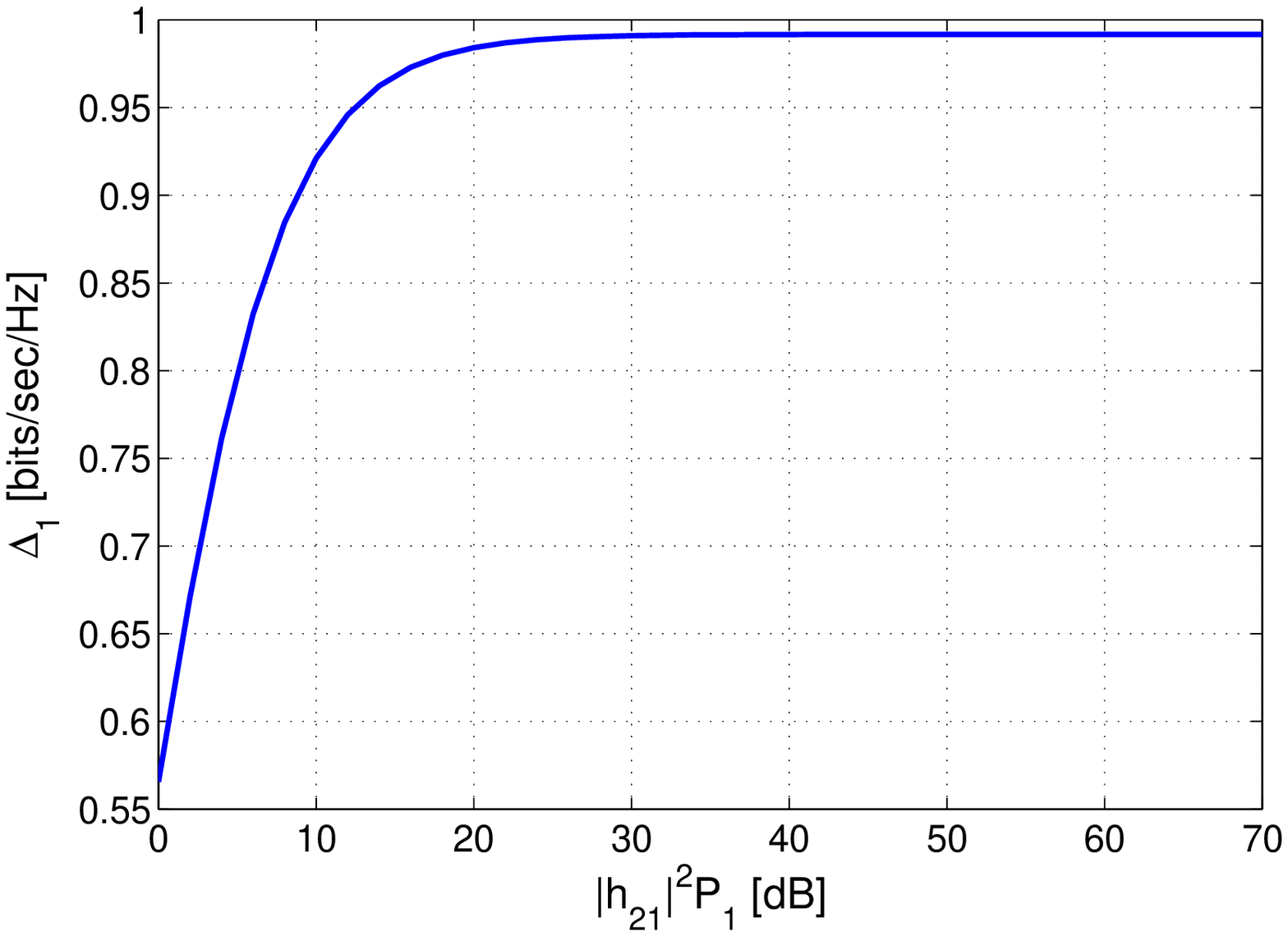}
\caption{$\Delta_1$ for various $|h_{21}|^2P_1$ values with $P_1=P_2$, $|h_{31}|^2=0.1|h_{21}|^2$ and $|h_{32}|^2=0.8|h_{21}|^2$.}
\label{Fig:Case2Results_CF}
\end{figure}

Fig. \ref{Fig:Case1Results_CF} shows $\Delta_1$ for various $|h_{21}|^2P_1$ values when $P_1=P_2$, $|h_{31}|^2=0.1|h_{21}|^2$ and $|h_{32}|^2=1.5|h_{21}|^2$.
This explains the gap $\Delta_1$ is always less than one bit for the case of $|h_{21}|^2P_1\leq |h_{32}|^2P_2$.
Fig. \ref{Fig:Case2Results_CF} shows $\Delta_1$ for various $|h_{21}|^2P_1$ values when $P_1=P_2$, $|h_{31}|^2=0.1|h_{21}|^2$ and $|h_{32}|^2=0.8|h_{21}|^2$.
In this case of $|h_{21}|^2P_1 > |h_{32}|^2P_2$,
the gap $\Delta_1$ becomes quite close to one as $|h_{21}|^2$ increases, but still less than one.


\section{Gaussian Relay Channel Capacity to Within Two Bits: AF Relaying Scheme} \label{Sec:TwoBits}
Although the DF and CF relaying schemes work well in the Gaussian relay channel,
both of them need a {\it smart} relay that can decode or compress the received signal and re-encode it.
In this section, we propose a {\it very} simple AF-based relaying scheme for a {\it dumb} relay
and show that it can achieve the Gaussian relay channel capacity to within $2$ bits/sec/Hz regardless of the channel parameters.

We first find an explicit expression for the achievable rate of the AF relaying scheme.
For simplicity, we use previously defined $a$, $b$ and $c$ with $\theta_a\triangleq\angle h_{31}$, $\theta_b\triangleq\angle h_{32}$ and $\theta_c\triangleq\angle h_{21}$.
After the relay amplifies the received signal $Y_{2,i-1}$ at time $i-1$ and forwards it to the destination at time $i$ as
\setlength{\arraycolsep}{0.0em}
\begin{eqnarray}
X_{2,i} &{}={}& \sqrt{\frac{P_2}{c+1}}Y_{2,i-1}    \nonumber \\
&{}={}& \sqrt{\frac{P_2}{c+1}} \left( \sqrt{\frac{c}{P_1}}e^{j\theta_c}X_{1,i-1}+Z_{2,i-1}  \right),
\end{eqnarray}
the destination receives $Y_{3,i}$ at time $i$ as
\setlength{\arraycolsep}{0.0em}
\begin{eqnarray}
Y_{3,i} &{}={}& \sqrt{\frac{a}{P_1}}e^{j\theta_a}X_{1,i}+\sqrt{\frac{b}{P_2}}e^{j\theta_b} X_{2,i}      \nonumber \\
&{}={}& \sqrt{a}e^{j\theta_a}\frac{X_{1,i}}{\sqrt{P_1}}+\sqrt{\frac{bc}{c+1}}e^{j(\theta_b+\theta_c)}\frac{X_{1,i-1}}{\sqrt{P_1}}    \nonumber \\
&&  + \sqrt{\frac{b}{c+1}}e^{j\theta_b}Z_{2,i-1}+ Z_{3,i},
\end{eqnarray}
where we assume $\mathbb E[|X_{1,i}|^2]=P_1$ and $\mathbb E[|X_{2,i}|^2]=P_2$.
For simplicity, we normalize noise power by dividing $Y_{3,i}$ by $\sqrt{\frac{b}{c+1}+1}$ as
\setlength{\arraycolsep}{0.0em}
\begin{eqnarray}
\tilde Y_{3,i} &{}\triangleq{}&  \frac{Y_{3,i}}{\sqrt{\frac{b}{c+1}+1}} = \sum_{k=0}^1 H_k\tilde X_{1,i-k}+ \tilde Z_{3,i},
\end{eqnarray}
where
\setlength{\arraycolsep}{0.0em}
\begin{eqnarray}
[H_0\quad H_1]  &{}\triangleq{}& \left[\sqrt{\frac{a(c+1)}{b+c+1}}e^{j\theta_a}\quad \sqrt{\frac{bc}{b+c+1}}e^{j(\theta_b+\theta_c)}\right],\quad\;\; \\
\tilde X_{1,i}&{}\triangleq{}&\frac{X_{1,i}}{\sqrt{P_1}}, \\
\tilde Z_{3,i}&{}\triangleq{}&\frac{1}{\sqrt{\frac{b}{c+1}+1}} \left( \sqrt{\frac{b}{c+1}}e^{j\theta_b}Z_{2,i-1} + Z_{3,i} \right)    \nonumber \\
&& \sim \mathcal{CN}(0,1).
\end{eqnarray}
Hence, the AF relying scheme turns the channel from the source to the destination into a unit-memory intersymbol interference channel
\cite{KramerGastparGupta:05,KramerMaricYates:07}.
Following \cite{HirtMassey:88,ChengVerdu:93}, the achievable rate for the AF relaying scheme is then written by
\setlength{\arraycolsep}{0.0em}
\begin{eqnarray}
\max_{\Sigma(w)\geq 0 \atop \frac{1}{2\pi}\int_0^{2\pi}\Sigma(w)dw\leq 1} \!\!\!\!\!\!\!\!\!
    \frac{1}{2\pi}\int_0^{2\pi}\log_2\left(1+ \Sigma(w) \left|H(w)\right|^2\right)dw   \qquad
\end{eqnarray}
where $H(w)$ is the Fourier transform of $H_k$ given by
\setlength{\arraycolsep}{0.0em}
\begin{eqnarray}
H(w) = \sqrt{\frac{a(c+1)}{b+c+1}}e^{j\theta_a} + \sqrt{\frac{bc}{b+c+1}}e^{j(\theta_b+\theta_c-w)},
\end{eqnarray}
and $\frac{1}{2\pi}\int_0^{2\pi}\Sigma(w)dw\leq 1$ is the normalized power constraint.
Although the optimal power allocation $\Sigma^*(w)$ is the well-known water-filling \cite{KramerMaricYates:07,HirtMassey:88,ChengVerdu:93},
we here assume uniform power allocation as $\Sigma(w)=1$ for all $w$.
Using
\setlength{\arraycolsep}{0.0em}
\begin{eqnarray}
\int_0^{2\pi} \ln(\mu+\nu\cdot\cos (x+y))dx = 2\pi\ln\frac{\mu+\sqrt{\mu^2-\nu^2}}{2},\quad
\end{eqnarray}
from $\int_0^\pi \ln(\mu+\nu\cdot\cos x)dx = \pi\ln\frac{\mu+\sqrt{\mu^2-\nu^2}}{2}$ for $\mu\geq\nu>0$ in \cite[p.526]{GradshteynRyzhik:00}, we obtain
\setlength{\arraycolsep}{0.0em}
\begin{eqnarray}
R_{AF} &{}={}& \frac{1}{2\pi}\int_0^{2\pi}\log_2\bigg( 1+\frac{a(c+1)}{b+c+1}+\frac{bc}{b+c+1} \nonumber \\
&& \quad + \frac{2\sqrt{abc(c+1)}}{b+c+1}\cos (w+\theta_a-\theta_b-\theta_c) \bigg) dw    \nonumber \\
&{}={}& \log_2\frac{K+\sqrt{L}}{1+b+c}-1,   \quad
\label{Eqn:R_AF}
\end{eqnarray}
where
\setlength{\arraycolsep}{0.0em}
\begin{eqnarray}
K &{}={}& 1+a+b+c+(a+b)c,     \label{Eqn:K} \\
L &{}={}& (1+c)\left\{ (1+a+b)^2 +\left((a-b)^2+2(a+b) +1 \right)c \right\}.    \nonumber \label{Eqn:L} \\
\end{eqnarray}

Then, $R_{AF}$ in (\ref{Eqn:R_AF}) can be rewritten as
\setlength{\arraycolsep}{0.0em}
\begin{eqnarray}
&&R_{AF} = -1 + \log_2\left(1+|h_{31}|^2P_1\right)   \nonumber \\
        &&\quad + \log_2\left( 1+\frac{|h_{32}|^2P_2\left(|h_{21}|^2-|h_{31}|^2\right)P_1 + \sqrt{L}}
            {\left(1+|h_{31}|^2P_1\right)\left(1+|h_{32}|^2P_2+|h_{21}|^2P_1 \right)} \right).   \nonumber \label{Eqn:R_AF2}\\
\end{eqnarray}
Interestingly, we can see that as long as $|h_{21}|^2> |h_{31}|^2$, (\ref{Eqn:R_AF2}) is similar to $C_d$ in (\ref{Eqn:DRelayChCapacity})
and $R_{CF}$ in (\ref{Eqn:ModifiedR_CF}) such that
$\log_2(1+|h_{31}|^2P_1)$ bits are directly delivered from the source to the destination while
the remaining bits are additionally delivered through the relaying path.
This makes us to conjecture that $C^+-R_{AF}\leq 2$ is satisfied for $|h_{21}|^2> |h_{31}|^2$ where
we use two bits instead of one bit since there is a penalty of $-1$ bit in (\ref{Eqn:R_AF2}).
For $|h_{21}|^2 \leq |h_{31}|^2$, the capacity can be achieved within one bit by simply ignoring the relay as in \cite{AvestimehrDiggaviTse:07}.
In this case, if the relay is active, then the signal-to-noise ratio (SNR) at the destination becomes worse than that for the inactive relay
since too much noise is amplified at the relay and forwarded to the destination.
Finally, we can expect the capacity is achieved within $2$ bits/sec/Hz regardless of the channel parameters.

Getting an insight from the above argument, we propose an AF-based relaying scheme as follows.
\begin{itemize}
\item If $|h_{31}|^2 > |h_{21}|^2$, the relay is ignored and the achievable rate is equal to $\log_2(1+|h_{31}|^2P_1)$.
In this case, it is easily shown that $C^+ - \log_2(1+a)\leq 1$ as in \cite{AvestimehrDiggaviTse:07} by
\setlength{\arraycolsep}{0.0em}
\begin{eqnarray}
C^+ \leq C_1^+(0) = \log_2(1+a+c) \leq \log_2(1+a)+1. \quad\;
\end{eqnarray}
\item If $|h_{31}|^2 \leq |h_{21}|^2$, the relay amplifies $Y_{2,i-1}$ and forwards it to the destination as
\begin{eqnarray}
X_{2,i}=\sqrt{\frac{P_2}{|h_{21}|^2P_1+1}}\;Y_{2,i-1},
\end{eqnarray}
where we assume $\mathbb E[|X_{1,i}|^2]=P_1$ and $\mathbb E[|X_{2,i}|^2]=P_2$. 
\end{itemize}
From now on, we show that $C^+-R_{AF}\leq 2$ in the case of $a \leq c$.
Especially, we first consider the case of $a\leq c\leq b$, and then the case of $a\leq c$ and $b<c$.

\subsection{The case of $a\leq c\leq b$ $\left(|h_{31}|^2P_1\leq |h_{21}|^2P_1\leq |h_{32}|^2P_2 \right)$}
Defining $\Delta_2\triangleq C^+ - R_{AF}$, we get
\setlength{\arraycolsep}{0.0em}
\begin{eqnarray}
\Delta_2 &{}={}& 1+ \log_2\frac{M}{K+\sqrt{L}},
\end{eqnarray}
where
\setlength{\arraycolsep}{0.0em}
\begin{eqnarray}
M &{}={}& (1+a+c)(1+b+c).
\end{eqnarray}
Since showing $\log_2\frac{M}{K+\sqrt{L}} \leq 1$ is equivalent to showing $2(K+\sqrt{L})-M\geq 0$,
we first consider
\setlength{\arraycolsep}{0.0em}
\begin{eqnarray}
&&2\left(K+\sqrt{L}\right) - M \geq 2\left(K+c(b-a) \right) - M  \nonumber \\
&&\qquad\qquad\quad = (1-a+3c)b + (1-c)(1+a+c),
\end{eqnarray}
and let
$g(b) \triangleq (1-a+3c)b + (1-c)(1+a+c)$.
From $(1-a+3c)\geq 0$, it is notable that if $g(c)\geq 0$, then $g(b)\geq 0$ for all $b\geq c$.
Since
\setlength{\arraycolsep}{0.0em}
\begin{eqnarray}
g(c) = 1+a+c+2c(c-a) > 0,
\end{eqnarray}
$\Delta_2\leq 2$ for all $a\leq c\leq b$.
It is also notable when $a=b=c\rightarrow\infty$, we get $\Delta_2\rightarrow 2$.

\subsection{The case of $a\leq c$ and $b<c$ $\big(|h_{31}|^2P_1 \leq |h_{21}|^2P_1$ and $|h_{32}|^2P_2 < |h_{21}|^2P_1$\big)}
From (\ref{Eqn:UB-Case2}) and (\ref{Eqn:R_AF})--(\ref{Eqn:L}), we get
\setlength{\arraycolsep}{0.0em}
\begin{eqnarray}
&&\Delta_2 = 1+ \log_2\frac{P+\sqrt{Q}}{(a+c)\left(K+\sqrt{L}\right)},
\end{eqnarray}
where
\setlength{\arraycolsep}{0.0em}
\begin{eqnarray}
P &{}={}& (1+b+c)\left((1+a+b)(a+c)-2ab\right),   \\
Q &{}={}& 4(1+b+c)^2abc(a-b+c).
\end{eqnarray}
Since
\setlength{\arraycolsep}{0.0em}
\begin{eqnarray}
2(a+c)K-P &{}={}&a^2(1-b+c)+a(1+b+c)^2 \nonumber \\
&&+c(1+(c-b)b+c)\geq 0,
\end{eqnarray}
showing $\log_2\frac{P+\sqrt{Q}}{(a+c)\left(K+\sqrt{L}\right)} \leq 1$ is equivalent to showing
$\big( 2(a+c)(K+\sqrt{L}) - P\big)^2 - Q  \geq 0$.
To do it, we first consider the case of $a\leq b< c$ and then the case of $b<a\leq c$.
\subsubsection{The case of $a\leq b< c$}
After some manipulation, we obtain
\setlength{\arraycolsep}{0.0em}
\begin{eqnarray}
&&\left( 2(a+c)(K+\sqrt{L}) - P\right)^2 - Q  \nonumber \\
&&\quad \geq \left( 2(a+c)(K+c(b-a)) - P\right)^2 - Q    \nonumber \\
&&\quad = (a+c)\left( \beta_0+ \beta_1c+\beta_2c^2+\beta_3c^3\right),
\end{eqnarray}
where
\setlength{\arraycolsep}{0.0em}
\begin{eqnarray}
\beta_0 &{}={}& a\big( a^2(5-2b+b^2) + (1+b)^2(5+2b+b^2) \nonumber \\
&& + 2a(5+5b-b^2-b^3) \big),   \\
\beta_1 &{}={}& 5 + 8 b + 2 b^2 + b^4 + 2 a^3 (3 + b) + a^2 (21 + 2 b - 11 b^2) \nonumber \\
&& + 4 a (5 + 8 b + 5 b^2 + 2 b^3),    \\
\beta_2 &{}={}& 10 + a^3 + a^2 (12 - 8 b) + 22 b + 6 b^2 - 6 b^3 \nonumber \\
&& + a (21 + 14 b + 13 b^2),    \\
\beta_3 &{}={}& 5 + 6 a + (a-b)^2 + 14 b + 8b(b-a).
\end{eqnarray}
Since $s(c)\triangleq\beta_0+ \beta_1c+\beta_2c^2+\beta_3c^3$ is a cubic function with $\beta_3 > 0$,
it is notable that if $s(b)\geq 0$, $s'(b)\geq 0$ and $s''(b)\geq 0$, then $s(c)\geq 0$ for all $c\geq b$.
From $b\geq a$, we get
\setlength{\arraycolsep}{0.0em}
\begin{eqnarray}
&&s(b) = a^3 (5 + 4 b) + a^2 (10 + 31 b + 12 b^2) \nonumber \\
&&\quad + b (5 + 18 b + 29 b^2 + 20 b^3) + a (5 + 32 b + 63 b^2 + 44 b^3) \nonumber \\
&&\quad + 4ab^2(a-b)^2+4b^3(b^2+2ab-3a^2) \geq 0,   \\
&&s'(b) = 5 + 6 a^3 + 28 b + 61 b^2 + 54 b^3 + a^2 (21 + 26 b) \nonumber \\
&&\quad + a (20 + 74 b + 66 b^2) + 4ab(a-b)^2 + 16b^2(b^2-a^2) \nonumber \\
&&\quad > 0,   \\
&&s''(b) = 20 + 24 a^2 + 74 b + 96 b^2 + a (42 + 64 b) \nonumber \\
&&\quad + 2a(a-b)^2 + 6b(7b^2-6ab-a^2) > 0.
\end{eqnarray}
Hence, we have $\Delta_2\leq 2$ for all $c> b\geq a$.

\subsubsection{The case of $b<a\leq c$}
Similarly, we obtain
\setlength{\arraycolsep}{0.0em}
\begin{eqnarray}
&&\left( 2(a+c)(K+\sqrt{L}) - P\right)^2 - Q  \nonumber \\
&&\quad \geq \left( 2(a+c)(K+c(a-b)) - P\right)^2 - Q    \nonumber \\
&&\quad = (a+c)\left( \gamma_0+ \gamma_1c+\gamma_2c^2+\gamma_3c^3\right),
\end{eqnarray}
where
\setlength{\arraycolsep}{0.0em}
\begin{eqnarray}
\gamma_0 &{}={}& a \big(a^2 (5 - 2 b + b^2) + (1 + b)^2 (5 + 2 b + b^2)     \nonumber \\
&& +2 a (5 + 5 b - b^2 - b^3)\big),    \\
\gamma_1 &{}={}& 5 + a^3 (14 - 6 b) + 8 b + 2 b^2 + b^4 + 4 a (5 + 6 b + b^2)   \nonumber \\
&& + a^2 (29 + 10 b + 5 b^2),   \\
\gamma_2 &{}={}& 28 a^2 + 9 a^3 + a (29 - 2 b - 11 b^2) \nonumber \\
&& + 2 (5 + 7 b + 3 b^2 + b^3), \\
\gamma_3 &{}={}& 5 + 14 a + (a-b)^2 + 6 b + 8a(a-b).
\end{eqnarray}
From $a\geq b$, $t(c)\triangleq\gamma_0+ \gamma_1c+\gamma_2c^2+\gamma_3c^3$ is a cubic function with $\gamma_3 > 0$.
Since
\setlength{\arraycolsep}{0.0em}
\begin{eqnarray}
&&t(a) = 2 a (5 + 28 a^3 + 10 b + 6 b^2 + 2 b^3 + a^2 (34 + 6 b)    \nonumber \\
&&\quad + 4 a (5 + 6 b + b^2)) + 2a(a^2-b^2)^2 + 16a^4(a-b)    \nonumber \\
&&\quad \geq 0,    \\
&&t'(a) = 5 + 112 a^3 + 8 b + 2 b^2 + 6 a^2 (17 + 4 b)      \nonumber \\
&&\quad + 4 a (10 + 13 b + 4 b^2) + 2b^2(a-b)^2     \nonumber \\
&&\quad + 37a^4 - 36a^3b - b^4  > 0, \\
&&t''(a) = 4 \big(5 + 35 a^2 + 7 b + 3 b^2 + a (22 + 8 b)\big) + 4b(a-b)^2 \nonumber \\
&&\quad + 8a(9a^2 - 8ab - b^2) > 0,
\end{eqnarray}
$t(c)\geq 0$ for all $c\geq a$.
Hence, $\Delta_2\leq 2$ for all $c\geq a > b$.
Finally, we conclude the proposed AF relaying scheme achieves the Gaussian relay channel capacity to within $2$ bits/sec/Hz regardless of the channel
parameters.
Moreover, when $a=b\rightarrow\infty$ and $c\rightarrow\infty$, we have $\Delta_2\rightarrow 2$.

\begin{figure}[h]
\centering
\includegraphics[width=3.5in]{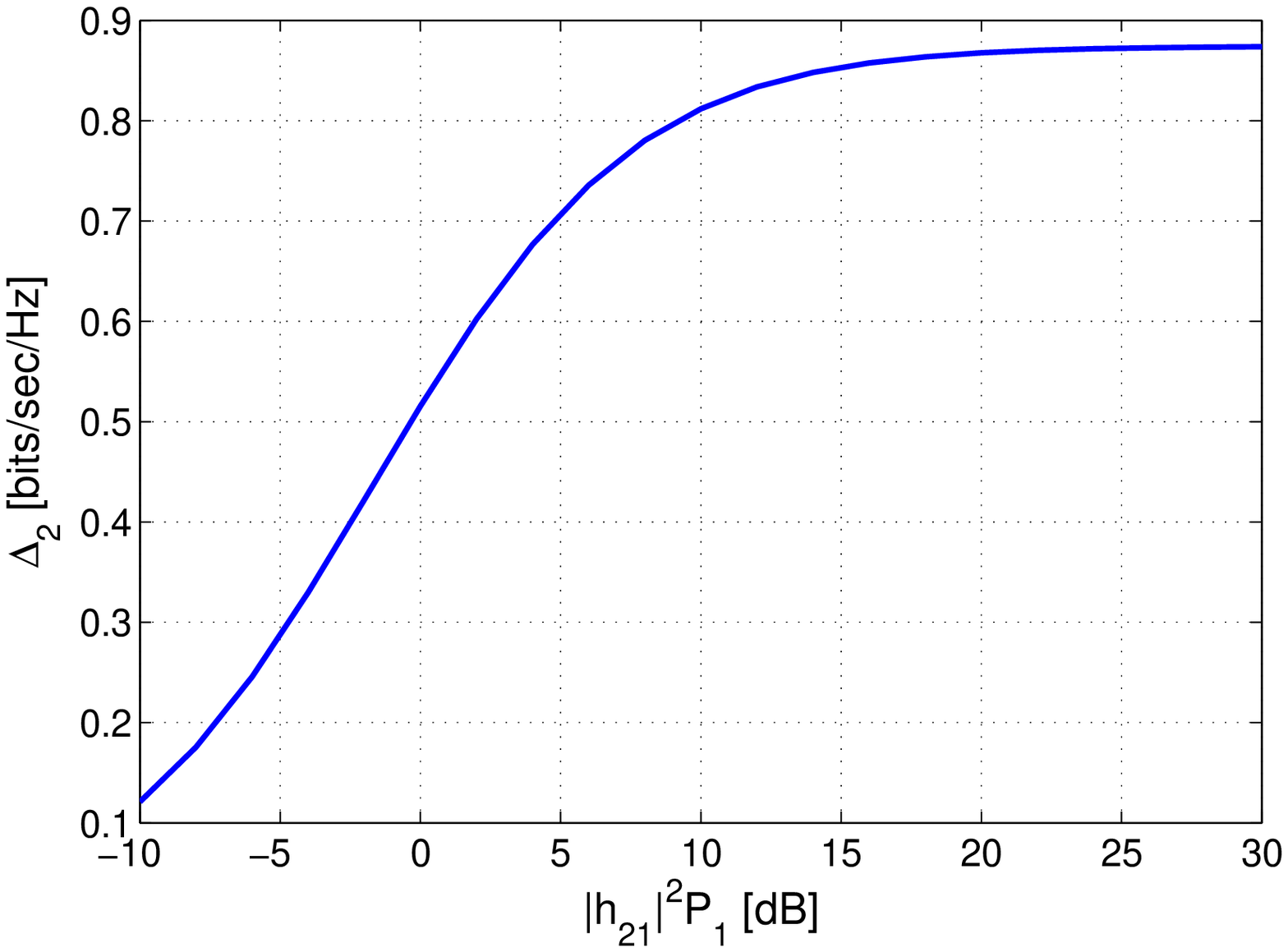}
\caption{$\Delta_2$ for various $|h_{21}|^2P_1$ values with $P_1=P_2$, $|h_{31}|^2=0.1|h_{21}|^2$ and $|h_{32}|^2=1.5|h_{21}|^2$.}
\label{Fig:Case1Results_AF}
\end{figure}

\begin{figure}[h]
\centering
\includegraphics[width=3.5in]{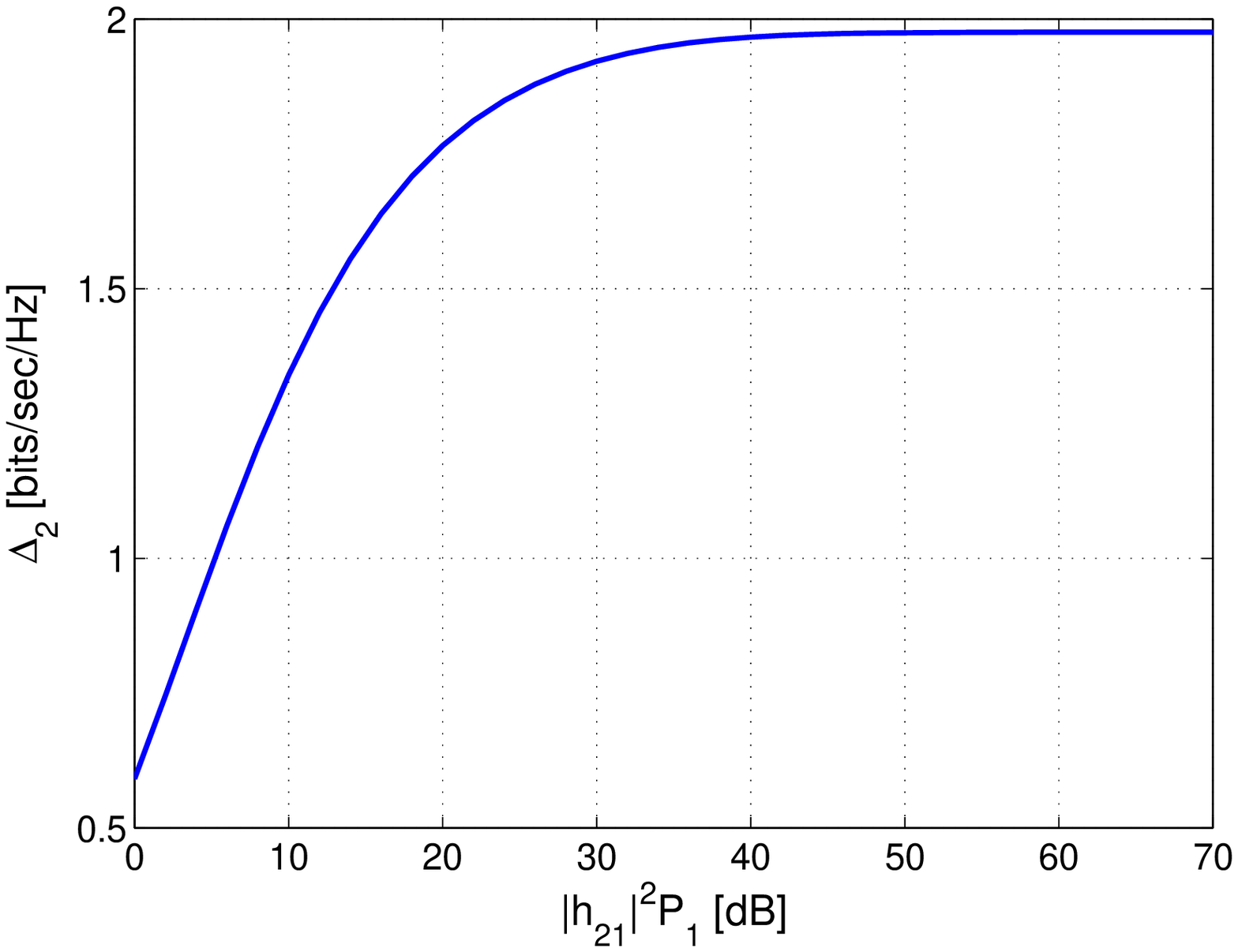}
\caption{$\Delta_2$ for various $|h_{21}|^2P_1$ values with $P_1=P_2$, $|h_{31}|^2=0.48|h_{21}|^2$ and $|h_{32}|^2=0.5|h_{21}|^2$.}
\label{Fig:Case2Results_AF}
\end{figure}

Fig. \ref{Fig:Case1Results_AF} shows $\Delta_2$ for various $|h_{21}|^2P_1$ values when $P_1=P_2$, $|h_{31}|^2=0.1|h_{21}|^2$ and $|h_{32}|^2=1.5|h_{21}|^2$.
This explains the gap $\Delta_2$ is always less than two bits for the case of $|h_{31}|^2P_1\leq |h_{21}|^2P_1\leq |h_{32}|^2P_2 $.
Fig. \ref{Fig:Case2Results_AF} shows $\Delta_2$ for various $|h_{21}|^2P_1$ values when $P_1=P_2$, $|h_{31}|^2=0.48|h_{21}|^2$ and $|h_{32}|^2=0.5|h_{21}|^2$.
In this case of $|h_{31}|^2P_1 \leq |h_{21}|^2P_1$ and $|h_{32}|^2P_2 < |h_{21}|^2P_1$, since $|h_{31}|^2P_1$ is quite close to $|h_{32}|P_2$,
the gap $\Delta_2$ also becomes quite close to two as $|h_{21}|^2$ increases, but still less than two.

\end{document}